\documentstyle[12pt]{article}
\begin{document}

\title{A General Classification of Three-Neutrino Models and $U_{e 3}$}

\author{{\bf S.M. Barr}\\ and \\{\bf Ilja Dorsner}\\ Bartol Research
Institute\\ University of Delaware\\ Newark, DE 19716}

\date{BA-00-15}
\maketitle

\begin{abstract}

A classification of models with three light neutrinos 
is given. This classification includes virtually
all of the three-neutrino models
models proposed in the last few years, of which there are
approximately one hundred. The essential
ideas, attractive features, and characteristic problems of
the different classes of model are discussed. The classification is
based principally on how large
$\nu_{\mu} - \nu_{\tau}$ mixing is obtained. A general discussion of 
the mixing parameter $U_{e 3}$ is then given, showing what values are
to be expected for it in each type of model.

\end{abstract}

\newpage

\section{Introduction}

Over the years, and especially since the discovery of the large
mixing of $\nu_{\mu}$ seen in atmospheric neutrino
experiments, there have been numerous models of neutrino masses
proposed in the literature. In the last two years alone, as
many as one hundred different models have been published.
One of the goals of this paper is to give a helpful classification
of these models. Such a classification is possible because
in actuality there are only a few basic ideas that underlie the
vast majority of published neutrino mixing schemes. 
After some preliminaries, we present in section 2 
a general classification of three-neutrino models that have a 
hierarchical neutrino spectrum. In section 3
we discuss the parameter $U_{e3}$, which describes the `1-3' mixing
of neutrinos. Since theoretical models are constructed to 
account for the solar and atmospheric neutrino oscillation data,
which tightly constrain the `1-2' and `2-3' mixings but not the `1-3'
mixing, the parameter $U_{e3}$ will be very important in the future
for distinguishing
among different kinds of models and testing particular schemes.

There are four indications of neutrino mass that have guided
recent attemps to build models: (1) the solar neutrino problem,
(2) the atmospheric neutrino anomaly, (3) the LSND experiment, and
(4) dark matter. There are many excellent reviews of the evidence for
neutrino mass.$^1$

(1) The three most promising solutions to the solar 
neutrino problem are based on neutrino mass. These are the
small-angle MSW solution (SMA), the large-angle MSW solution (LMA), and 
the vacuum oscillation solution (VO). All these solutions involve
$\nu_e$ oscillating into some other type of neutrino ---
in the models we shall consider, predominantly $\nu_{\mu}$. In the
SMA solution the mixing angle and mass-squared splitting between
$\nu_e$ and the neutrino into which it oscillates are roughly
$\sin^2 2 \theta \sim 5.5 \times 10^{-3}$ and $\delta m^2 
\sim 5.1 \times 10^{-6} eV^2$. 
For the
LMA solution one has $\sin^2 2 \theta \sim 0.79$, and $\delta m^2
\sim 3.6 \times 10^{-5} eV^2$. (The numbers are best-fit
values from a recent analysis.$^2$) And for the VO solution 
$\sin^2 2 \theta \sim 0.93$, and 
$\delta m^2 \sim 4.4 \times 10^{-10} eV^2$. (Again, these are best-fit
values from a recent analysis.$^3$)

(2) The atmospheric neutrino anomaly strongly implies that $\nu_{\mu}$
is oscillating with nearly maximal angle into either $\nu_{\tau}$
or a sterile neutrino, with the data preferring the former
possibility.$^4$ One has $\sin^2 2 \theta \sim 1.0$, and $\delta m^2
\sim 3 \times 10^{-3} eV^2$.

(3) The LSND result, which would indicate a mixing between $\nu_e$
and $\nu_{\mu}$ with $\delta m^2 \sim 0.3 - 1 eV^2$ is regarded with more 
skepticism for two reasons.
The experimental reason is that KARMEN has failed to corroborate
the discovery, although it is true that KARMEN has not excluded the
whole LSND region.
The theoretical reason is that to account for
the LSND result and also for both the solar and atmospheric anomalies
by neutrino oscillations would require three quite different
mass-squared splittings, and that can only be achieved with
{\it four} species of neutrino. This significantly complicates
the problem of model-building. In particular, it is regarded as
not very natural, in general, to have a fourth sterile neutrino
that is extremely light compared to the weak scale. (There are some
theoretical frameworks that can give light sterile particles, but
they tend to give many of them, not just one.)
For these
reasons, we assume that
the LSND results do not need to be explained by neutrino
oscillations, and the classification we present
includes only three-neutrino models.

(4) The fourth possible indication of neutrino mass is the existence of
dark matter. If a significant amount of this dark matter is in
neutrino mass, it would imply a neutrino mass of order several eVs .
In order then to achieve the small mass splittings needed to
explain the solar and atmospheric anomalies one would have to assume 
that $\nu_e$, $\nu_{\mu}$ and $\nu_{\tau}$ were nearly degenerate.
We shall not focus on such models in our classification, which
is primarily devoted to models with ``hierarchical" neutrino masses. 
However, in most models with nearly degenerate masses, the neutrino mass
matrix consists of a dominant piece proportional to the identity
matrix and a much smaller hierarchical piece. Since the oscillations
are caused by the small hierarchical piece, such
models can be classified together with hierarchical models.

In sum, the models we shall classify are those which assume
(a) three flavors of neutrino that oscillate ($\nu_e$, $\nu_{\mu}$,
and $\nu_{\tau}$), (b) 
the atmospheric anomaly explained by $\nu_{\mu}$-$\nu_{\tau}$
oscillations with nearly maximal angle, and (c) the solar
anomalies explained by $\nu_e$ oscillating primarily with $\nu_{\mu}$
with either small angle (SMA) or large angle (LMA, VO).

There are several major divisions among models. One is
between models in which the neutrino masses arise through
the see-saw mechanism,$^5$ and those in which the neutrino masses are
generated directly at low energy. In see-saw models, there are
both left- and right-handed neutrinos. Consequently, there are
five fermion mass matrices to explain: the four Dirac mass
matrices, $U$, $D$, $L$, and $N$ of the up quarks, down quarks, 
charged leptons, and neutrinos, respectively, and the Majorana
mass matrix $M_R$ of the right-handed neutrinos. The four Dirac
mass matrices are all roughly of the weak scale, while $M_R$
is generally much larger than the weak scale. After
integrating out the superheavy right-handed neutrinos,
the mass matrix of the left-handed neutrinos is given by
$M_{\nu} = - N^T M_R^{-1} N$. In conventional see-saw models
there are three right-handed neutrinos, one for each of the three
families of quarks and leptons. And, typically, in such conventional 
see-saw models there is a close relationship between the $3 \times 3$
Dirac mass matrix $N$ of the neutrinos and the other $3 \times 3$ Dirac
mass matrices $L$, $U$, and $D$. Usually the four Dirac matrices are
related to each other by grand unification and/or flavor 
symmetries.
That means that in conventional see-saw models neutrino masses and mixings  
are just one aspect of the larger problem of quark and lepton masses, and
are likely to shed great light on that problem, and perhaps even be
the key to solving it. However, in most see-saw models the Majorana
matrix $M_R$ is either not related or is tenuously related to the
Dirac mass matrices of the quarks and leptons. The freedom in
$M_R$ is the major obstacle to making precise predictions of
neutrino masses and mixings in most see-saw schemes.

There are also what we shall refer to as unconventional see-saw models
in which the fermions that play the role of the heavy
right-handed partners of the neutrinos are not part of the ordinary
three-family structure but are some other neutral fields. There need not
be three of them, and the Dirac mass matrix of the neutrinos therefore
need not be $3 \times 3$ nor need it have any particular connection
to the other Dirac mass matrices $L$, $U$, and $D$. Such unconventional
see-saw models we classify together with non-see-saw models.

In non-see-saw schemes, there are no right-handed neutrinos.
Consequently, there are only four mass matrices to
consider, the Dirac mass matrices of the quarks and charged leptons,
$U$, $D$, and $L$, and the Majorana mass matrix of the light
left-handed neutrinos $M_{\nu}$. Typically in such schemes $M_{\nu}$
has nothing directly to do with the matrices $U$, $D$, and $L$, but
is generated at low-energy by completely different physics.

The three most popular possibilities in recent models for generating
$M_{\nu}$ at low energy in a non-see-saw way are (a) triplet Higgs,
(b) variants of the Zee model,$^6$ and (c) R-parity violating terms in
low-energy supersymmetry. (a) In triplet-Higgs models, $M_{\nu}$ arises
from a renormalizable term of the form $\lambda_{ij} \nu_i \nu_j H_T^0$,
where $H_T$ is a Higgs field in the $(1,3, +1)$ representation
of $SU(3) \times SU(2) \times U(1)$. (b) In the Zee model, the Standard
Model
is supplemented with a scalar, $h$, in the $(1,1, +1)$ representation
and having weak-scale mass. 
This field can couple to the lepton doublets $L_i$ as $L_i L_j h$
and to the Higgs doublets $\phi_a$ (if there is more than one)
as $\phi_a \phi_b h$. Clearly it is not possible to assign
a lepton number to $h$ in such a way as to conserve it in both 
these terms. The
resulting lepton-number violation allows one-loop diagrams that 
generate a Majorana mass for the left-handed neutrinos. (c) In
supersymmetry the presence of such R-parity-violating terms in
the superpotential as $L_i L_j E^c_k$ and $Q_i D^c_j L_k$,
causes lepton-number violation, and allows one-loop diagrams that
give neutrino masses. Neutrino mass can also arise at tree level from 
R-parity-violating terms of the form $H_u L_i$,
which mix neutrino and Higgs superfields
and lead to sneutrino vacuum expectation values.

It is clear that in all of these schemes the couplings that give
rise to neutrino masses have little to do with the physics that
gives mass to the other quarks and leptons. While this allows
more freedom to the neutrino masses, it would from one point of view
be very disappointing, as it would mean that the observation of
neutrino oscillations is almost irrelevant to the great question
of the origin of quark and charged lepton masses.

It should also be mentioned that some models derive the neutrino mass
matrix $M_{\nu}$ directly from non-renormalizable terms 
of the form $\nu_i \nu_j H_u H_u/M$ 
without specifying where these terms come from. While such terms 
do arise in the conventional see-saw mechanism
they can also arise in other ways. Models in which these operators
do not arise from a see-saw or where their origin is left
unspecified we classify as non-see-saw models.

Another major division among models has to do with the kinds of symmetries
that constrain the forms of mass matrices and that, in some models,
relate different mass matrices to each other. There are two
main approaches: (a) grand unification, and (b) flavor symmetry. Many
models use both. 

(a) The simplest grand unified group is $SU(5)$.
In minimal $SU(5)$ there is one relation among the Dirac mass matrices,
namely $D= L^T$, coming from the fact that the left-handed charged
leptons are unified with the right-handed down quarks in a 
$\overline{{\bf 5}}$, while the right-handed charged leptons and 
left-handed down quarks are unified in a ${\bf 10}$. In $SU(5)$
there do not have to be right-handed neutrinos, though they may
be introduced. In $SO(10)$, which in several ways is a very
attractive group for unification, the minimal model gives the relations
$N = U  \propto D = L$. In realistic models these relations
are modified in various ways, for example by the appearance
of Clebsch coefficients in certain entries of some of the mass matrices.
It is clear that unified symmetries are so powerful that very
predictive models are possible. Most of the published models which give
sharp predictions for masses and mixings are unified models.

(b) Flavor symmetries can be either abelian or non-abelian.
Non-abelian symmetries are useful for obtaining the 
equality of certain elements of the mass matrix, as in models
where the neutrino masses are nearly degenerate, and in the
so-called ``flavor democracy" schemes, which will be discussed later.
Abelian symmetries
are useful for explaining hierarchical mass matrices through the
so-called Froggatt-Nielson mechanism.$^7$ The idea is simply
that different elements of the mass matrices arise at different
orders in flavor symmetry breaking. In particular,
different fermion multiplets can differ in charge under a $U(1)$ flavor 
symmetry that is spontaneously broken by some ``flavon" 
expectation value (or values), $\langle f_i \rangle$. Thus,
different elements of the fermion mass matrices would be suppressed by
different powers of $\langle f_i \rangle/M \equiv \epsilon_i \ll 1$,
where $M$ is the scale of flavor physics. This kind of
scheme can explain small mass ratios and mixings in the sense of
predicting them to arise at certain orders in the small quantities 
$\epsilon_i$. A drawback of such models compared to many
grand unified models is that actual numerical predictions,
as opposed to order of magnitude estimates, are not possible.
On the other hand, models based on flavor symmetry involve less
of a theoretical superstructure built on top of the Standard Model
than do unified models, and could therefore be considered more
economical in a certain sense. Unified models put more in but get more
out than do abelian-flavor-symmetry models.

The most significant new fact about neutrino mixing is the largeness
of the mixing between $\nu_{\mu}$ and $\nu_{\tau}$
This comes as somewhat of a surprise from the point of view of both
grand unification and flavor symmetry approaches. Since grand
unification relates leptons to quarks, one might expect lepton 
mixing angles to be small like those of the quarks. In particular,
the mixing between the second and third family of quarks is given by
$V_{cb}$, which is known to be $0.04$. That is to be
compared to the nearly maximal mixing of the second and third
families of leptons: $U_{\mu 3} \cong 1/\sqrt{2}
\cong 0.7$. It is true that even in the early 1980's some grand unified
models predicted large neutrino mixing angles. (Especially
noteworthy is the remarkably prophetic 1982 paper of Harvey, Ramond, and 
Reiss,$^8$
which explicitly
predicted and emphasized that there should be
large $\nu_{\mu}- \nu_{\tau}$ mixing. However, 
in those days the top mass was expected to be 
light, and that paper assumed it to be 25 GeV. That gave $V_{cb}$ 
to be about $0.22$. The corresponding lepton mixing
was further boosted by a Clebsch of 3. With the actual value of $m_t$
that we now know, the model of Ref. 8 would predict $U_{\mu 3}$ to
be only 0.12). What makes the largeness of $U_{\mu 3}$ a puzzle
in the present situation is the fact that we now know that both
$V_{cb}$ and $m_c/m_t$ are exceedingly small.

The same puzzle exists in the context of flavor symmetry. The fact
that the quark mixing angles are small suggests that there is a
family symmetry that is only weakly broken, while the large mixings
of some of the neutrinos would suggest that family symmetries are badly
broken.

The first point of interest, therefore, in looking at any model of neutrino
mixing is how it explains the large mixing of $\nu_{\mu}$ and 
$\nu_{\tau}$. This will be the feature that we will use to organize
the classification of models.

\section{Classification of three-neutrino models}

Virtually all three-neutrino models published in the last few years
fit somewhere in the simple classification now to be described.
In fact, almost all of them are cited below.
The main divisions of this classification are based on how the
large $\nu_{\mu}-\nu_{\tau}$ mixing arises. This mixing is
described by the element $U_{\mu 3} \equiv \sin \theta_{23}$ 
of the so-called MNS matrix$^9$
(analogous to the CKM matrix for the quarks).

The mixing angles of the neutrinos are the mismatch between the
eigenstates of the neutrinos and those of the charged leptons, or
in other words between the mass matrices $L$ and $M_{\nu}$.
Thus, there are two obvious ways of obtaining large $\theta_{23}$:
either $M_{\nu}$ has large off-diagonal elements while $L$ is
nearly diagonal, or $L$ has large off-diagonal elements and $M_{\nu}$
is nearly diagonal. Of course this distinction only makes sense
in some preferred basis. But in almost every model there is some 
preferred basis given by the underlying symmetries of that model.
This distinction gives the first major division in the classification,
between models of what we shall call class I and class II. 
(It is also possible that the
large mixing is due almost equally to large off-diagonal elements in
$L$ and $M_{\nu}$, but this possibility seems to be realized
in very few published models. We will put them into class II.)

If the large $\theta_{23}$ is due to $M_{\nu}$ (class I), then it
becomes important whether $M_{\nu}$ arises from a non-see-saw
mechanism or the see-saw mechanism. We therefore distinguish
these cases as class I(1) and class I(2) respectively. In the
see-saw models, $M_{\nu}$ is given by $- N^T M_R^{-1} N$, so
a further subdivision is possible: models in which the large mixing
comes from large off-diagonal elements in $M_R$ we call I(2A); models
in which the large mixing comes from large off-diagonal elements in $N$
we call I(2B); and models in which neither $M_R$ nor $N$ have large 
off-diagonal elements but $M_{\nu}= - N^T M_R^{-1} N$ nevertheless does
we call I(2C).

The other main class of models, where
$\theta_{23}$ is due to large off-diagonal elements in $L$ the mass matrix
of the {\it charged} leptons, we called class II. The question in these 
models is why, given that $L$ has large off-diagonal elements, there
are not also large off-diagonal elements in the Dirac mass matrices
of the other charged fermions, especially $D$ (which is typically
closely related to $L$), causing
large CKM mixing of the quarks. In the literature there seem
to be two ways of answering this question. One way involves
the CKM angles being small due to a cancellation between large angles
that are nearly equal in the up and down quark sectors. We call
this class II(1). The main
examples of this idea are the so-called ``flavor democracy models".
The other idea is that the matrices $L$ and $D^T$ (related by
unified or flavor symmetry) are ``lopsided" in such a way that
the large off-diagonal elements only affect the mixing of fermions
of one handedness: left-handed for the leptons, making $U_{\mu 3}$
large, and right-handed for the quarks, leaving $V_{cb}$ small.
We call this approach class II(2).

Schematically, one then has

\begin{equation}
\begin{array}{ll}
I & {\rm Large \; mixing \; from } \; M_{\nu}  \\
& (1) \;\; {\rm Non \; see \; saw} \\
& (2) \;\; {\rm See \; saw} \\
& \;\;\;\;\; {\rm A.  \; Large \; mixing \; from } \; M_R \\
& \;\;\;\;\; {\rm B.  \; Large \; mixing \; from } \; N \\
& \;\;\;\;\; {\rm C.  \; Large \; mixing \; from } \; - N^T M_R^{-1} N \\
II & {\rm Large \; mixing \; from } \; L  \\
& (1) \;\; {\rm CKM \; small \; by \; cancellation} \\ 
& (2) \;\; {\rm lopsided } \; L.
\end{array}
\end{equation}

Now let us examine the different categories in more detail, 
giving examples from the literature.

\vspace{0.2cm}

\noindent
{\bf I(1) Large mixing from $M_{\nu}$, non-see-saw}. 

This kind of model gives a natural explanation of the discrepancy
between the largeness of $U_{\mu 3} = \sin \theta_{23}$ and 
the smallness of $V_{cb}$.
$V_{cb}$ comes from Dirac mass matrices, which are all presumably
nearly diagonal like $L$, whereas $U_{\mu 3}$ comes from the matrix 
$M_{\nu}$; and since in non-see-saw models $M_{\nu}$ comes from  
completely different physics than do the Dirac mass matrices
it is not at all surprising that it has a very different form
from the others, containing some large off-diagonal elements.
While this basic idea is very simple and appealing, these models
have the drawback that in non-see-saw models the
form of $M_{\nu}$, since it comes from new physics unrelated to
the origin of the other mass matrices, is highly unconstrained.
Thus, there are few definite predictions, in general, for masses
and mixings in such schemes. However, in some schemes constraints
can be put on the new physics responsible for
$M_{\nu}$. 

As we saw, there are a variety of attractive ideas for generating
a non-see-saw $M_{\nu}$ at low energy, and there are published
models of neutrino mixing corresponding to all these 
ideas.$^{10-21}$
$M_{\nu}$ comes from triplet Higgs in Refs. 10-12; from 
the Zee mechanism in Refs. 13-15; 
and from R-parity and lepton-number-violating
terms in supersymmetry in Refs. 16 and 17. 

In Ref. 18 a ``democratic form"
of $M_{\nu}$ is enforced by a family symmetry. (The democratic form is
one in which all the elements of the matrix are equal or very nearly
equal. In most schemes of
``flavor democracy", as we shall see later, it is the charged
lepton mass matrix $L$ that is assumed to have a democratic form and
$M_{\nu}$ is assumed approximately diagonal, giving models
of class II(1), But in Ref. 18 the
opposite is assumed.)
Several other models
in class I(1) exist in the literature.$^{19,20}$

There is a basic question that has to be answered by any model
of class I(1), namely why the mass splitting seen in solar neutrino
oscillations ($\delta m^2_{12}$) is much smaller than that seen in
atmospheric oscillations ($\delta m^2_{23}$). If all the elements of
$M_{\nu}$ were of the same order, then indeed large mixing angles
would be typical, as desired to explain the atmospheric neutrino
oscillations, but the neutrino mass ratios would then also be typically
of order unity, and one would expect $\delta m^2_{12} \sim \delta
m^2_{23}$. Conversely, if there is a small parameter in
$M_{\nu}$ that accounts for the ratios of mass splittings, then
the question arises why the mixing angles are not also controlled by that
small parameter.

A satisfactory answer to these questions requires that 
$M_{\nu}$ have a special form. Three satisfactory forms are
possible, as has been pointed out in several analyses.$^{21}$
We shall consider them in turn.

(a) In the literature one finds that the majority of models of class I(1)
(and, as we shall see later, many models of other classes too) 
assume the following form for $M_{\nu}$:

\begin{equation}
M_{\nu} = \left( \begin{array}{ccc}
m_{11} & m_{12} & m_{13} \\ m_{12} & s^2 M + O(\delta) M  & sc M +
O(\delta) M \\ m_{13} & sc M + O(\delta) M & c^2 M + O(\delta) M \end{array}
\right),
\end{equation}

\noindent
where $s \equiv \sin \theta$, $c \equiv \cos \theta$, $\theta \sim \pi/4$,
$\delta \ll 1$, and $m_{ij} \ll M$.
By a rotation in the
2-3 plane by an angle close to $\theta$, the 2-3 block will be diagonalized
and the matrix will take the form

\begin{equation}
M_{\nu}' \cong \left( \begin{array}{ccc}
m_{11} & c m_{12} - s m_{13} & c m_{13} + s m_{12} \\
c m_{12} - s m_{13} & O(\delta) M & 0 \\ c m_{13} + s m_{12} &
0 & M 
\end{array} \right).
\end{equation}

\noindent
It is clear that for $m_{ij} \stackrel{_<}{_\sim} 
\delta M$ there is a hierarchy of mass
eigenvalues, and that $\delta m^2_{23} \cong M^2$ and 
$\delta m^2_{12} = O(\delta^2) M^2$. On the other hand the
atmospheric neutrino angle $\theta_{23}$, which is approximately
given by $\theta$, is of order one.
The value of the solar angle depends on the size of $m_{ij}$.
In particular $\theta_{12} \sim (c m_{12} - s m_{13})/
(\delta M)$. Consequently, either small angle or large angle 
solutions of the solar neutrino problem can be naturally obtained.

One sees that in order to get the hierarchy among neutrino mass
splittings and at the same time a large atmospheric angle
one has assumed a form for $M_{\nu}$ in Eq. (2) that has 
a special relationship among the 22, 23, 32, and 33
elements. If such a relationship existed simply
accidentally, then the model would be ``fine-tuned" to some extent.
Specifically, the 2-3 block of $M_{\nu}$ would have a determinant
that was of $O(\delta)$ times its ``natural" value. 

A number of the models in the literature that are of class I(1)
are indeed fine-tuned in this way. However, two ways of achieving the
special form in Eq. (2) in a technically natural way in models of class I(1)
have been proposed in the literature: (i) factorization, and (ii) 
permutation symmetry. 

(i) The idea of factorization is that 
the neutrino mass matrix arises from one-loop contributions that
are dominated by a single diagram, giving $(M_{\nu})_{ij}
\cong \lambda_i \lambda_j M$, where $\lambda_i$ is the coupling
of $\nu_i$ to the particles in the loop. If $\lambda_2 \sim
\lambda_3$ then the form in Eq. (2) results. A good example of this kind
of model is Ref. 16, where $\lambda_i$ is an R-parity-violating
and lepton-number-violating coupling of $\nu_i$ to a quark-squark
(or lepton-slepton) pair in supersymmetry. 

A factorized form of
$M_{\nu}$ can also arise at tree-level by a non-standard see-saw mechanism
in which $\lambda_i$ is a Dirac coupling of $\nu_i$ to a {\it single}
heavy Majorana fermion that is integrated out. This is the
basic idea in the papers in Ref. 20. (A special case of this
is supersymmetric models with R-parity-violating terms that mix neutrinos
with other neutralinos. In these the neutralinos play the role
of the heavy fermions in the see-saw, and factorized forms of $M_{\nu}$
can result.)

(ii) The other idea for achieving the form in Eq. (2) is permutation symmetry.
The basic idea is to use non-abelian symmetry to relate different
elements of $M_{\nu}$. Generally the relationship will be one
of equality, thus giving maximal mixing angles.
A good example is the model of Ref. 10, in which an $S_2 \times S_2$ 
permutation symmetry among
four left-handed neutrinos is used to obtain the form 

\begin{equation}
M_{\nu} = \left( \begin{array}{cccc} A & B & C & D \\
B & A & D & C \\ C & D & A & B \\ D & C & B & A 
\end{array} \right).
\end{equation}

\noindent
Then by assuming that the linear combination of $(\nu_1 - \nu_2)/\sqrt{2}$
acquires a superlarge Majorana
mass, the residual three light species of neutrino
end up with a mass matrix

\begin{equation}
M_{\nu}' = \left( \begin{array}{ccc}
A+B & F & F \\ F & A & B \\ F & B & A \end{array} \right)
= (A+B) I + \left( \begin{array}{ccc} 0 & F & F \\
F & - B & B \\ F & B & - B \end{array} \right), 
\end{equation}

\noindent
which in effect has the form in Eq. (2), since the part proportional
to the identity does not contribute to oscillations. From this one
sees that it is possible to get the form in Eq. (2) in a technically
natural way using flavor symmetries. Again, 
either small or large solar angle can arise depending on the magnitude of
$F/B$.

(b) Another form for $M_{\nu}$ that is satisfactory is

\begin{equation}
M_{\nu} = \left( \begin{array}{ccc} 
m_{11} & c M & s M \\ c M & m_{22} & m_{23} \\
s M & m_{23} & m_{33} \end{array} \right),
\end{equation}

\noindent
where $m_{ij} \ll M$, and as before $s \equiv \sin \theta$ and 
$c \equiv \cos \theta$, with $\theta \sim \pi/4$.
By a rotation in the 2-3 plane by angle $\theta$ one brings this
to the form 

\begin{equation}
M_{\nu}' = \left( \begin{array}{ccc}
m_{11} & M & 0 \\ M & m_{22}' & m_{23}' \\
0 & m_{23}' & m_{33}' \end{array} \right).
\end{equation}

\noindent
This pseudo-Dirac form in the 1-2 block shows that $\nu_1$ and
$\nu_2$ will be maximally mixed with nearly degenerate masses
approximately equal to $M$, while the third neutrino will have
smaller mass. Thus $\delta m^2_{23} \cong M^2$ and $\delta m^2_{12}
\sim m_{ij} M$. Such a form always gives bimaximal mixing, i.e.
large mixing angle for both atmospheric neutrinos and solar
neutrinos, in contrast to Eq. (2) which can give either large
or small angle solutions for the solar neutrino problem.

The form in Eq. (6) can easily be achieved using various family
symmetries.$^{11, 12, 15}$ A particularly interesting possibility$^{12,15}$
is that the symmetry in question is $L_e - L_{\mu} - L_{\tau}$, which
if exact would allow only the large elements of order $M$ in Eq. (6).

An interesting and instructive model in which $M_{\nu}$ is of the form
given in Eq. (6) is found in Ref. 13. This model is
based on the Zee mechanism, which gives
a neutrino mass matrix $M_{\nu}$ that is symmetric but has vanishing diagonal 
elements. In particular it can give a matrix of the form

\begin{equation}
M_{\nu} \cong \left( \begin{array}{ccc}
0 & m/\sqrt{2} & -m/\sqrt{2} \\ m/\sqrt{2} & 0 & \Delta \\
-m/\sqrt{2} & \Delta & 0 \end{array} \right),
\end{equation}

\noindent
where $\Delta \ll m$. 
There is some mild fine-tuning in this model in the sense that in order for
the 12 and 13 elements of $M_{\nu}$ to be nearly equal in magnitude
(as must be
so to have nearly maximal atmospheric angle) a relation among the
couplings and masses of the Zee model must be satisfied that has no
basis in symmetry.

(c) A third possible form for $M_{\nu}$ is

\begin{equation}
M_{\nu} = \left( \begin{array}{ccc} M' & m_{12} & m_{13} \\
m_{12} & m_{22} & M \\ m_{13} & M & m_{33} \end{array} \right),
\end{equation}

\noindent
where $m_{ij} \ll M$.
In such a scheme, $\nu_{\mu}$ and $\nu_{\tau}$ are nearly maximally
mixed and nearly degenerate, with $\delta m^2_{23} \sim m_{ij} M 
\ll M^2$. Therefore, in order for the splitting $\delta m^2_{12}$ to
be even smaller, it must be that $M' \cong M$ to great accuracy. 
If this is not to be a fine-tuning of parameters, then it must be
the consequence of some non-abelian flavor symmetry.

\vspace{0.2cm}

\noindent
{\bf I(2A) See-saw $M_{\nu}$, large mixing from $M_R$}

In models of class of I(2), as in class I(1), the large atmospheric
neutrino mixing angle comes from $M_{\nu}$, which however
is now assumed to arise from the conventional see-saw 
mechanism. $M_{\nu}$ therefore
has the form $- N^T M_R^{-1} N$, where $N$ is a $3 \times 3$
matrix typically related by symmetry to $L$, $U$, and $D$. In class I(2A), 
the large off-diagonal elements in
$M_{\nu}$ are assumed to come from $M_R$, while the Dirac neutrino
matrix $N$ is assumed to be nearly diagonal and hierarchical like the other
Dirac matrices $L$, $U$, and $D$. Many examples of models of class I(2A)
exist in the literature.$^{22-30}$
As with the models of class I(1),
these models have the virtue of explaining in a natural way
the difference between
the lepton angle $U_{\mu 3}$ and the quark angle $V_{cb}$. The
quark mixings all come from Dirac matrices, while the lepton mixings
involve the Majorana matrix $M_R$, which it is quite reasonable to
suppose might have a very different character, with large off-diagonal
elements.

However, there is a general problem with models of this type, which
not all the examples in the literature convincingly overcome.
The problem is that if $N$ has a hierarchical and nearly diagonal
form, it tends to communicate this property to $M_{\nu}$. For example,
suppose we take $N = {\rm diag}( \epsilon', \epsilon, 1) M$, with
$1 \gg \epsilon \gg \epsilon'$. And suppose that the $ij^{th}$
element of $M_R^{-1}$ is called $a_{ij}$. Then the matrix
$M_{\nu}$ will have the form

\begin{equation}
M_{\nu} \propto \left( \begin{array}{ccc}
\epsilon^{\prime2} a_{11} & \epsilon' \epsilon a_{12} &
\epsilon' a_{13} \\ \epsilon' \epsilon a_{12} & \epsilon^2 a_{22} &
\epsilon a_{23} \\ \epsilon' a_{13} & \epsilon a_{23} & 
a_{33} \end{array} \right). 
\end{equation}

\noindent
If all the non-vanishing elements $a_{ij}$ are of the same
order of magnitude, then obviously $M_{\nu}$ is approximately
diagonal and hierarchical. 
The contribution to the leptonic angles coming from $M_{\nu}$ would
therefore typically be proportional to the small parameters $\epsilon$
and $\epsilon'$. One way that a $\theta_{23}$ of $O(1)$ could
arise is that
the small parameter coming from $N$ gets
cancelled by a correspondingly large parameter from $M_R^{-1}$.$^{22}$
The trouble is that to have such a relationship between the magnitudes
of parameters in $N$ and $M_R$ is usually unnatural, since these
matrices have very different origins. This problem has been 
pointed out by various authors.$^{23}$ We shall call it the
Dirac-Majorana conspiracy problem.

This problem is avoided in models in which the hierarchies in 
$N$ and $M_R$ are controlled by the same family symmetries and the
same small parameters. Example of such correlated hierarchies can
be found in the papers of Ref. 24.

Another way of getting around the Dirac-Majorana conspiracy problem
is to assume a special form for $M_R$. An apparently simple solution
is to take the 2-3 block of $M_R$ to be skew diagonal.
For example, suppose

\begin{equation}
M_R \cong \left( \begin{array}{ccc} M' & 0 & 0 \\ 
0 & 0 & M \\ 0 & M & 0 
\end{array} \right), \;\;\; N \cong \left( \begin{array}{ccc}
\epsilon' & & \\ & \epsilon & \\ & & 1 \end{array} \right).
\end{equation}

\noindent
where $\epsilon' \ll \epsilon \ll 1$. 
Then the 2-3 block of $M_{\nu} = - N^T M_R^{-1} N$ is also approximately
skew diagonal, and one has that $\nu_{\mu}$ and $\nu_{\tau}$ are
nearly degenerate and maximally mixed, as needed to explain the
atmospheric neutrino anomaly. A number of models in the literature
exploit this idea.$^{25}$

Unfortunately, as most of the papers in Ref. 25 noted, this idea
has a problem with solar neutrinos. The problem is that
it is unnatural in such a scheme for the splitting $\delta m^2_{12}$ to
be smaller than $\delta m^2_{23}$.
One has $m_2 \cong m_3 \cong \epsilon M$ and $m_1 \cong 
\epsilon^{\prime 2} M'$. Therefore, unless $M'$ is tuned with great accuracy
this scheme cannot give a satisfactory solution to the solar neutrino
problem.

It is clear that if one seeks to avoid the Dirac-Majorana conspiracy
problem and also to explain both solar 
and atmospheric neutrino oscillations, 
an even cleverer choice of
the forms of $N$ and $M_R$ must be found. Several papers have found
such forms.$^{26-29}$. In the model of Ref. 27, for instance, 
the Dirac and Majorana matrices of the neutrinos have the forms

\begin{equation}
N = \left( \begin{array}{ccc} x^2 y & 0 & 0 \\ 0 & x & x \\
0 & O(x^2) & 1 \end{array} \right) m_D, \;\;\;
M_R = \left( \begin{array}{ccc} 0 & 0 & A \\ 0 & 1 & 0 \\
A & 0 & 0 \end{array} \right) m_R,
\end{equation}

\noindent
where $x$ and $y$ are small parameters. If one computes $M_{\nu}
= -  N^T M_R^{-1} N$ one finds that

\begin{equation}
M_{\nu} = - \left( \begin{array}{ccc} 0 & O(x^4 y/A) & x^2 y/A \\
O(x^4 y/A) & x^2 & x^2 \\ x^2 y/A & x^2 & x^2 \end{array} \right)
m_D^2/m_R.
\end{equation}

\noindent
Observe that this gives a maximal mixing of the second and third families,
without having to assume any special relationship between the
small parameters in $N$ (namely $x$, and $y$) and the parameter in $M_R$
(namely $A$). This example is generalized in the papers of Ref. 28.

Note that the matrix in Eq. (13) is of the general form given in
Eq. (2), but here it arises through the see-saw mechanism.
An interesting point about the form of $M_{\nu}$ in Eq. (13) is that
it gives bimaximal mixing. This is easily seen by doing a
rotation of $\pi/4$ in the 2-3 plane, bringing the matrix to
the form 

\begin{equation}
M_{\nu}' = \left( \begin{array}{ccc} 0 & z & z' \\ z & 0 & 0 \\
z' & 0 & 2 x^2 \end{array} \right).
\end{equation}

\noindent
In the 1-2 block this matrix has a Dirac form, giving nearly maximal
mixing of $\nu_e$.

Another interesting model that avoids the Dirac-Majorana conspiracy
problem, but requires a mild fine-tuning to get the hierarchy among
neutrino mass splittings, is given in Ref. 29. The Majorana and Dirac
neutrino mass matrices in that model have the form

\begin{equation}
N = \left( \begin{array}{ccc} 0 & 0 & x \\ 0 & x & 0 \\ x & 0 & 1
\end{array} \right) m, \;\;\;\; M_R^{-1} = \left( \begin{array}{ccc}
a M^{-1} & b M^{-1} & 0 \\ b M^{-1} & c M^{-1} & 0 \\ 0 & 0 & M^{\prime -1}
\end{array} \right),
\end{equation}

\noindent
where $x \ll 1$, $M/M' \ll x^2$, and $a,b,c \sim 1$. This gives

\begin{equation}
M_{\nu} = \left( \begin{array}{ccc} \epsilon x^2 & 0 & \epsilon x \\
0 & c & b \\ \epsilon x & b & a + \epsilon \end{array} \right) 
(m^2 x^2/M).
\end{equation}

\noindent
Here $\epsilon \equiv \frac{M}{M'} \frac{1}{x^2} \ll 1$. The 
atmospheric neutrino angle will be of order unity if $a$, $b$, and
$c$ are all of the same order, which requires no fine-tuning or
Dirac-Majorana conspiracy. However, to make $\delta m^2_{12}
\ll \delta m^2_{23}$ requires that the condition $\sqrt{ac - b^2}
\ll a, b, c$, which does not arise from any symmetry,
be satisfied.

Models of class I(2A) can be constructed that 
predict either small or large values of
the solar neutrino angle $\theta_{12}$.

\newpage

\noindent
{\bf I(2B) See-saw $M_{\nu}$, large mixing from $N$}

We now turn to see-saw models in which the large atmospheric neutrino angle
comes from large off-diagonal elements in the Dirac neutrino mass matrix 
$N$ rather than in the Majorana matrix $M_R$.

At least at first glance, this seems to be a less natural approach.
The point is that
if the large $\theta_{23}$ is due to large off-diagonal elements in $N$,
it might be expected that the other Dirac mass matrices, $U$, $D$, and $L$,
would also have large off-diagonal elements, giving large CKM angles.
The model in Ref. 31 only attempts to describe the lepton sector and   
so does not resolve this problem. In Ref. 32 it is assumed that $N$ has
large off-diagonal elements and $L$ does not, but the difference in
character of these matrices is not accounted for. In the interesting model
of Ref. 33 the difference between $N$ and the other Dirac matrices is
accounted for by a fine-tuning.
In that model
all of the quark and lepton mass matrices are given (in terms of relatively
few parameters) by linear combinations of certain matrices
that are hierarchical and nearly diagonal. In order that $N$ have 
off-diagonal elements that are comparable to its diagonal elements,
an accidental cancellation must occur that suppresses the diagonal elements.

There are ways to construct models of class I(2B) in which the
difference between $N$ and the other Dirac matrices is
explained without fine-tuning.$^{34}$
However, 
experience seems to show that this approach is harder to make work
than the others. 

\vspace{0.2cm}

\noindent
{\bf I(2C) See-saw $M_{\nu}$, large mixing from $ - N^T M_R^{-1} N$}

In order for the see-saw mass matrix $M_{\nu} = -N^T M_R^{-1} N$
to have large off-diagonal elements it is not necessary that
either $M_R$ or $N$ have large off-diagonal elements, as emphasized in
Ref. 35. Following Ref. 35, consider the matrices 

\begin{equation}
N \sim \left( \begin{array}{ccc} \epsilon' & \epsilon' & \epsilon' \\
\epsilon' & \epsilon & \epsilon \\ \epsilon' & \epsilon & 1 
\end{array} \right) m, \;\;\; M_R^{-1} = \left( \begin{array}{ccc}
r_1 & 0 & 0 \\ 0 & r_2 & 0 \\ 0 & 0 & r_3 \end{array} \right) M^{-1},
\end{equation}

\noindent
where it is assumed that $r_2 \epsilon^2 \gg r_3, r_1 \epsilon^{\prime 2}$.
Then to leading order in small quantities $M_{\nu}$ has the form

\begin{equation}
M_{\nu} \sim \left( \begin{array}{ccc} (\epsilon'/\epsilon)^2 &
\epsilon'/\epsilon & \epsilon'/\epsilon \\ \epsilon'/\epsilon &
1 & 1 \\ \epsilon'/\epsilon' & 1 & 1 \end{array} \right) r_2 \epsilon^2
(m^2/M).
\end{equation}

\noindent
It is easy to understand what is happening. The fact that $r_2$ is
much larger than $r_1$ and $r_3$ means that the right-handed neutrino
of the second family is much lighter than the other two. Effectively,
then, one right-handed neutrino dominates $M_R^{-1}$. As a consequence
one obtains an approximately ``factorized" form for $M_{\nu}$, 
just as one did in the unconventional 
see-saw models considered in the papers of Ref. 20, in which a single
right-handed fermion also dominated. Those unconventional see-saw models
could also be considered as examples of class I(2C).

\vspace{0.2cm}

\noindent
{\bf II(1) Large mixing from $L$, CKM small by cancellation}
 
We now turn to those models in which the large value of $\theta_{23}$
comes predominantly from the charged lepton mass matrix $L$, with
$M_{\nu}$ being nearly diagonal. The issue that arises in such models
is whether the other Dirac mass matrices, especially $D$ and $U$,
also have large off-diagonal elements, and if so why this does
not lead to large CKM angles for the quarks. Some published models
do not deal with this question since they are only models of
the lepton sector and do not attempt to describe the quarks at all.$^{36}$
However, while it may in non-see-saw models make sense to discuss
$M_{\nu}$ apart from the other mass matrices, it would seem that
under most reasonable assumptions the matrix $L$ should have
some relationship to $U$ and $D$. 

Why, then,
are the CKM angles small? One possibility is that the CKM
angles are small because of an almost exact cancellation between
large angles needed to diagonalize $U$ and $D$. That, in turn,
would imply that $U$ and $D$, even though highly
non-diagonal, have nearly identical forms. This idea is realized in
most so-called ``flavor democracy" models.$^{37}$ 

A ``flavor-democratic" mass matrix is one in which all the elements are
equal:

\begin{equation}
M_{FD} \equiv \left( \begin{array}{ccc}
1 & 1 & 1 \\ 1 & 1 & 1 \\ 1 & 1 & 1 \end{array} \right) m_{\ell}
\end{equation}

\noindent
A Dirac mass matrix can have such a form
as the result of separate $S_3$ permutation
symmetries acting on the left-handed and right-handed fermions.
This form is of
rank 1, and thus gives only one family a mass. Of course, in
realistic models based on flavor democracy it is assumed that
the mass matrices also get small corrections that come from the
breaking of the permutation symmetries, which give rise to
masses for the lighter families. 

It is clear that the flavor democratic form is diagonalized by
a unitary matrix with rotation angles that are large. In fact,
the matrix is

\begin{equation}
U_{FD} = \left( \begin{array}{ccc} 1/\sqrt{2} & 1/\sqrt{6} & 1/\sqrt{3} \\
- 1/\sqrt{2} & 1/\sqrt{6} & 1/\sqrt{3} \\ 0 & -2/\sqrt{6}
& 1/\sqrt{3} \end{array} \right).
\end{equation}

\noindent
The reason why the CKM angles are small in flavor democracy models
is that both $U$ and $D$ are assumed to have approximately
the democratic form. Thus, the large rotation angles nearly cancel
between the up and down sectors. However, as first noted in Ref. 38, 
it is possible to have large neutrino mixing angles if $M_{\nu}$
is assumed to have not the democratic form but a nearly diagonal
form. This difference in form is plausible, given that $M_{\nu}$ is
a Majorana matrix rather than a Dirac matrix like the others.
In this case, the angles required to diagonalize $M_{\nu}$
would be small, and the MNS matrix would come predominantly from
diagonalizing $L$.

In Ref. 38, it is assumed that $M_{\nu}$ is diagonal and hierarchical
in form, and the elements of $M_{\nu}$ are assumed to arise entirely
from the breaking of the permutation symmetries of the model.
The model of Ref. 38 therefore has the neutrino masses being
hierarchical. However, most published
versions of the flavor democracy idea$^{39}$ assume
that $M_{\nu}$ is approximately proportional to the identity matrix.
The form $M_{\nu} \propto I$
is invariant under an $S_3$ permutation of the left-handed neutrinos.
(However, it should be noted that $M_{\nu} \propto I$ is not the
most general form consistent with permutation symmetry, and so to make
this form technically natural some further symmetries must be
invoked.)
Small deviations from the identity matrix would arise from
terms that break the flavor symmetries of the model. In such
versions, the three neutrino masses are nearly degenerate, but
the splittings can be made hierarchical to accomodate the solar
and atmospheric data.

An exact flavor-democratic form of
$L$ would leave two charged leptons degenerate and therefore one of the
neutrino mixing angles undefined. And if $M_{\nu}$ were
exactly proportional to the identity matrix, all three
neutrinos would be degenerate and all three neutrino mixing angles would
be undefined. 
Exactly what angles are predicted for the neutrinos depends, therefore,
on the form of the small 
contributions to the mass matrices
that break the permutation symmetries. There are many possibilities.
In some, the MNS matrix comes out to be very close to 
$U_{FD}^{\dag}$. However, it is not surprising, given the degeneracies
the the exact permutation-symmetric forms give, that the small 
permutation-symmetry-breaking contributions to the mass matrices can
lead to additional large mixings, and to forms for the MNS matrix that
depart significantly from $U_{FD}^{\dag}$.
It is typical in flavor democracy models for the 
angle $\theta_{12}$ to come out large, and in many cases it comes out
to be close to $\pi/4$ as in the matrix $U_{FD}^{\dag}$. However, it is
possible for $\theta_{12}$ to be small. This can happen if the
matrix $M_{\nu}$ is such that the neutrinos $\nu_1$ and $\nu_2$
form a pseudo-Dirac pair. Then the 1-2 angles from
the diagonalization of both $L$ and $M_{\nu}$ will be close to 
$\pi/4$ and their difference can be small.

The number of possible models, based on different ways to break
permutation symmetry,
is large.
There exists an extensive and growing literature in this area.
There are also many models based not on the pure flavor-democratic
form in Eq. (19), but on forms in which all the elements of the mass matrix
are assumed to be approximately equal in magnitude, but allowed
to differ in complex phase. This is sometimes called the  
``Universal Strength for Yukawa couplings" approach or USY.$^{40}$

The idea of flavor democracy is an elegant one, especially in
that it uses one basic idea to explain the largeness of
the leptonic angles, the smallness of the quark angles, and the
fact that one family is much heavier than the others. On the
other hand, it is based on very special forms for the mass matrices 
which come from very specific symmetries. It is in this sense 
a narrower approach to the problem of fermion masses than 
some of the others we have discussed. 

It would be interesting to know whether simple models of class II(1),
in which the CKM angles are small by cancellations of large
angles, can be constructed using ideas other than flavor
democracy.

\vspace{0.2cm}

\noindent
{\bf II(2) Large mixing from ``lopsided" $L$}

We now come to an idea for explaining the
largeness of $\theta_{23}$ that has great flexibility, in the
sense that it can be implemented in many different kinds of
models: grand unified, models with abelian or non-abelian flavor
symmetries, see-saw or non-see-saw neutrino masses, and so on.
The basic idea of the ``lopsided" $L$ approach
is that the charged-lepton and down-quark
mass matrices have the approximate forms

\begin{equation}
L \sim \left( \begin{array}{ccc} 0 & 0 & 0 \\ 0 & 0 & \epsilon \\
0 & \sigma & 1 \end{array} \right) m_D, \;\;
D \sim \left( \begin{array}{ccc} 0 & 0 & 0 \\ 0 & 0 & \sigma \\
0 & \epsilon & 1 \end{array} \right) m_D. 
\end{equation}

\noindent
The ``$\sim$" sign is used because in realistic models these 
$\sigma$ and $\epsilon$ entries could have additional factors
of order unity, such as from Clebsch coefficients. The fact that $L$
is related closely in form to the {\it transpose} of $D$ is
a very natural feature from the point of view of $SU(5)$
or related symmetries, and
is a crucial ingredient in this approach. 
The assumption is that $\epsilon \ll 1$, while $\sigma \sim 1$.
In the case of the charged leptons $\epsilon$ controls the
mixing of the second and third families of {\it right}-handed
fermions (which is not observable at low energies), while
$\sigma$ controls the mixing of the second and third families
of {\it left}-handed fermions, which contributes to $\theta_{23}$
and makes it large. For the quarks the reverse is the case
because of the ``$SU(5)$" feature: the small $O(\epsilon)$
mixing is in the left-handed sector, accounting for the smallness
of $V_{cb}$, while the large $O(\sigma)$ mixing is in the right-handed
sector, where it cannot be observed and does no harm.

In this approach the three crucial elements are these:
(a) Large mixing of neutrinos (in particular of $\nu_{\mu}$
and $\nu_{\tau}$) caused by large off-diagonal elements in
the {\it charged}-lepton mass matrix $L$; (b) these off-diagonal
elements appearing in a highly asymmetric or lopsided way; and
(c) $L$ being similar to the {\it transpose} of $D$ by $SU(5)$
or a related symmetry.

What makes this approach so flexible is that the problem of obtaining
a realistic pattern of neutrino masses
is decoupled from the problem of getting large $\theta_{23}$. The 
large $\theta_{23}$ arises from $L$ while the neutrino mass spectrum
arises from $M_{\nu}$. Thus one is freed from having very special
textures for $M_{\nu}$ as was the case in class I models. This is also
true of the flavor democracy schemes; however, there the necessity of
near cancellation between up and down quark angles forced a
very particular kind of mass matrix texture and flavor symmetry.
The lopsided mass matrices, by contrast, can be achieved in many
ways, as can be seen from Refs. 41-47.

The first paper that has all three elements that define this
approach seems to be Ref. 41, which proposed 
a very specific idea for generating the fermion
mass hierarchies. The ideas of that paper were further explored in Ref. 42.

The lopsided $L$ idea next is seen in three papers that appeared  
almost simultaneously.$^{43-45}$
It is interesting that the same basic mechanism of lopsided $L$ 
was arrived at
independently by these three groups of authors from completely different
starting points. In Ref. 43 the model is based on 
$E_7/SU(5)\times U(1)$, and the structure of the mass matrices is 
determined by the 
Froggatt-Nielson mechanism. In Ref. 44 the 
model is based on $SO(10)$, and does not use the
Froggett-Nielson approach. Rather, the constraints on the form
of the mass matrices come from assuming a ``minimal" set of Higgs
for $SO(10)$-breaking and choosing the smallest and simplest set of
Yukawa operators that can give realistic mass matrices for the quarks and
charged leptons. Though both Refs. 43 and 44
assume a unified symmetry larger than $SU(5)$, in both
it is the $SU(5)$ subgroup that plays the critical role in relating
$L$ to $D^T$. The model of Ref. 45,
like that of Ref. 43, uses the
Froggatt-Nielson idea, but is not based on a grand unified
group. Rather, the fact that $L$ is related to $D^T$ follows
ultimately from the requirement of anomaly cancellation for the
various $U(1)$ flavor symmetries of the model. However, it is well known
that anomaly cancellation typically enforces charge assignments that
can be embedded in unified groups. So that even though the model does
not contain an explicit $SU(5)$, it could be said to be ``$SU(5)$-like".

In Ref. 46 are listed numerous papers that have used the lopsided $L$
approach in the context of grand unified theories. A variety of
symmetries --- abelian, non-abelian continuous, and non-abelian discrete
--- are used in these models to constrain the forms of mass matrices.
In Ref. 47 are papers that are not unified and do not discuss the
quark mass matrices, so that the third element of the approach ($L$
being related to $D^T$ by a symmetry related to $SU(5)$) is not explicitly
present.

As pointed out in Ref. 48, models based on lopsided $L$ can give
either large-angle or small-angle solutions to the solar neutrino
problem.

\vspace{0.2cm}

\noindent
{\bf A predictive model with lopsided L}

\vspace{0.2cm}

We shall now briefly describe a particular model of class II(2).
A remarkable fact about this model is that it was not constructed to
explain neutrino phenomenology; rather it emerged from
the attempt to find a realistic model of the masses of the
charged leptons and quarks in the context of $SO(10)$.
In fact, it is one of the most predictive models of quark and
lepton masses that exists in the literature.
The idea of the model was to take the Higgs sector of $SO(10)$ to be as
minimal as possible, and then to find what this implied for
the mass matrices of the quarks and charged leptons. In fact, in the first
paper proposing this model$^{49}$ no attention was paid to the neutrino
mixings at all. Only subsequently was it noticed
that the model actually
predicts a large mixing of $\nu_{\mu}$ with $\nu_{\tau}$ and
this led to a second paper, in which the implications for
neutrino phenomenology were stressed.$^{44}$ The reason for the large
mixing of $\nu_{\mu}$ and $\nu_{\tau}$ in this model is 
precisely the fact that
the charged lepton mass matrix has a lopsided form.

The reason this lopsided form was built into the model 
of Refs. 44 and 49,
was that it
was necessary to account for certain well-known features of the
mass spectrum of the quarks. In particular, the mass matrix entry
that is denoted $\sigma$ in Eq. (21) above plays three crucial
roles in this model that have nothing to do with neutrino
mixing. (1) It is required to get the Georgi-Jarlskog$^{50}$ factor
of 3 between $m_{\mu}$ and $m_s$. (2) It explains the value of
$V_{cb}$. (3) It explains why $m_c/m_t \ll m_s/m_b$. Remarkably,
it turns out not only to perform these three tasks, but also gives
mixing of order 1 between $\nu_{\mu}$ and $\nu_{\tau}$. Not often
are four birds killed with one stone.

In constructing the model, several considerations played a part.
First, a ``minimal" set of Higgs for $SO(10)$ was assumed. It
has been shown$^{51}$ that the smallest set of Higgs that will allow
a realistic breaking of $SO(10)$ down to $SU(3) \times SU(2) \times U(1)$,
with natural doublet-triplet splitting,$^{52}$ consists of a single
adjoint (${\bf 45}$), two pairs of spinors (${\bf 16} + \overline{{\bf 16}}$),
a pair of vectors (${\bf 10}$), and some singlets. The adjoint, in
order to give the doublet-triplet splitting, must have a VEV proportional
to the $SO(10)$ generator $B-L$. This fact is an important constraint.
Second, it was required that the qualitative features of the quark and lepton
spectrum should not arise by artificial cancellations or numerical
accidents. Third, it was required that the Georgi-Jarlskog factor arise
in a simple and natural way. Fourth, it was required that the entries
in the mass matrices should come from operators of low dimension that
arise in simple ways from integrating out small representations
of fermions.

Having imposed these conditions of economy and naturalness, a structure
emerged that had just six effective Yukawa terms
(just five if $m_u$ is allowed to vanish). These gave the following
mass matrices:

\begin{equation}
\begin{array}{ll}
U^0 = \left( \begin{array}{ccc} \eta & 0 & 0 \\ 0 & 0 & \frac{1}{3}
\epsilon \\ 0 & - \frac{1}{3} \epsilon & 1 \end{array} \right) m_U, \;\;\;
& D^0 = \left( \begin{array}{ccc} 0 & \delta & \delta' \\ \delta &
0 & \sigma + \frac{1}{3} \epsilon \\ \delta' & - \frac{1}{3} \epsilon 
& 1 \end{array} \right) m_D \\
& \\
N^0 = \left( \begin{array}{ccc} \eta & 0 & 0 \\ 0 & 0 & - \epsilon \\
0 & \epsilon & 1 \end{array} \right) m_U, \;\;\; & L^0 = \left( 
\begin{array}{ccc} 0 & \delta & \delta' \\ \delta & 0 & - \epsilon \\
\delta' & \sigma + \epsilon & 1 \end{array} \right) m_D.
\end{array}
\end{equation}

\noindent
(The first papers$^{49,44}$ gave only the structures of the second and third
families, while this was extended to the first family in a subsequent 
paper.$^{53}$)
Here $\sigma \cong 1.8$, $\epsilon \cong 0.14$, $\delta \cong |\delta'|
\cong 0.008$, $\eta \cong 0.6 \times 10^{-5}$.
The patterns that are evident in these matrices are due to the
$SO(10)$ group-theoretical characteristics of the various Yukawa terms.
Notice several facts about the crucial parameter $\sigma$ that is
responsible for the lopsidedness of $L$ and $D$. First, if $\sigma$
were not present, then instead of the Georgi-Jarlskog factor of 3,
the ratio $m_{\mu}/m_s$ would be given by 9. (That is, the Clebsch
of $\frac{1}{3}$ that appears in $D$ due to the generator $B-L$
gets squared in computing $m_s$.) Since the large entry $\sigma$ overpowers
the small entries of order $\epsilon$, the correct Georgi-Jarlskog
factor emerges. Second, if $\sigma$ were not present, $U$ and
$D$ would be proportional, as far as the two heavier families are
concerned, and $V_{cb}$ would vanish. Third, by having $\sigma \sim
1$ one ends up with $V_{cb}$ and $m_s/m_b$ being of the same order 
($O(\epsilon)$) as is indeed observed. And since $\sigma$ does not appear
in $U$ (for group-theoretical reasons) the ratio $m_c/m_t$ comes out
much smaller, of $O(\epsilon^2)$, also as observed. In fact,
with this structure, the mass of charm is predicted correctly to within
the level of the uncertainties. 

Thus, for several reasons {\it that have nothing to do with neutrinos} one
is led naturally to exactly the lopsided form that is found to give an
elegant explanation of the mixing seen in atmospheric neutrino data.

From the very small number of Yukawa terms, and from the fact that
$SO(10)$ symmetry gives the normalizations of these terms, and not merely
order of magnitude estimates for them, it is not surprising
that many precise predictions result. In fact there are
altogether nine predictions.$^{53}$ Some of these are post-dictions
(including the highly non-trivial one for $m_c$). But several
predictions will allow the model to be tested in the future,
including predictions for $V_{ub}$, and the mixing angles $U_{e2}$
$U_{e3}$.

\section{Expectations for the parameter $U_{e3}$}

All of the models that we have discussed aim to explain the atmospheric
neutrino anomaly by saying that there is maximal mixing between
$\nu_{\mu}$ and $\nu_{\tau}$, i.e. that $U_{\mu 3} \cong 1/\sqrt{2}$,
and they all aim to explain the solar neutrino problem either
by the small-angle MSW solution, in which $U_{e2} \simeq 0.05$,
or by one of the large-angle solutions (large-angle MSW or vacuum
oscillation), in which $U_{e2} \simeq 1/\sqrt{2}$. In this section
we examine the other
mixing, which is described by $U_{e 3}$. $U_{e3}$ is independent 
of the other two mixings, and {\it a priori} could take values
ranging from zero up to the present limit of about 0.2 . However,
what we find is that 
the great majority of viable models give one of four
mixing patterns, which we label with the Greek letters $\alpha$ 
through $\delta$.
These patterns are indicated in the following table.

\vspace{0.2cm}

$$\begin{array}{l|c|l}
Pattern & U_{e2} \;\;\;\; & U_{e3} \\ \hline
& & \\
\alpha & \sin \theta_{LA} & O(m_{\nu_{\mu}}/m_{\nu_{\tau}}), \;\;
O(\sqrt{m_e/m_{\mu}}) \sim 0.05 \\ 
& & \\ 
\alpha' & \sin \theta_{LA} & \sqrt{m_e/m_{\mu}} \sin \theta_{atm}\\
& & \\
\alpha^{\prime \prime} 
& \sin \theta_{LA} & \frac{2}{\sqrt{6}} \sqrt{m_e/m_{\mu}} \\ \hline
& & \\
\beta & \sin \theta_{LA} & 0 \\ \hline
& & \\
\gamma & \sin \theta_{SA} & \sin \theta_{SA} \tan \theta_{atm} \\
& & \\ \gamma' & \sqrt{m_e/m_{\mu}} \cos \theta_{atm} & 
\sqrt{m_e/m_{\mu}} \sin \theta_{atm} 
\\ \hline
& & \\
\delta & \sin \theta_{SA} & \ll \sin \theta_{SA} \tan \theta_{atm}
\end{array}
$$

\vspace{0.5cm}

In this table $\theta_{SA}$ and $\theta_{LA}$ stand for the
value of $\theta_{12}$ in the small-angle and large-angle solutions
of the solar neutrino problem, respectively.
What one sees is that if the solar angle is maximal one expects 
either that $U_{e3}$ will be of order 0.05 (pattern $\alpha$) or
that it will vanish (pattern $\beta$). In most models that fit 
pattern $\alpha$ only the order of magnitude of $U_{e3}$ is predicted.
However, some models predict it sharply. A particularly interesting
prediction that arises in certain types of models is that 
$| U_{e3}| = \sqrt{m_e/m_{\mu}} \sin \theta_{atm}$, which
we distinguish with the name $\alpha'$. A special case of this
occurs in some flavor democracy models, where $\sin \theta_{atm} =
2/\sqrt{6}$, and we call this $\alpha^{\prime \prime}$.

If the solar angle is small, i.e. the small-angle MSW solution, one typically
finds one of two results for $U_{e3}$: either it is given by
the relation $U_{e3} \cong U_{e2} \tan \theta_{atm}$ (pattern
$\gamma$) or it is small compared to that value (pattern $\delta$).
In certain models with pattern $\gamma$ 
there is the further prediction
for the solar angle that $U_{e2} \cong \sqrt{m_e/m_{\mu}} \cos \theta_{atm}$.
We call this pattern $\gamma'$, and it leads to the same
prediction for $U_{e3}$ that one has 
in pattern $\alpha'$.

We will first give some general preliminaries and then proceed to analyze
different kinds of models, showing why the four patterns we have
described are the ones that arise in the great majority of published
models.

The lepton mixing matrix, or ``MNS matrix" has the form 

\begin{equation}
U_{MNS} = U_L^{\dag} U_{\nu},
\end{equation}

\noindent
where $U_L$ is the unitary matrix that diagonalizes $L^{\dag} L$,
and $U_{\nu}$ is the unitary matrix that diagonalizes 
$M_{\nu}^{\dag} M_{\nu}$. It is convenient to write $U_L$
in the form

\begin{equation}
U_L = \left( \begin{array}{ccc}
1 & 0 & 0 \\ 0 & \overline{c}_{23} & \overline{s}_{23} \\
0 & - \overline{s}_{23} & \overline{c}_{23} \end{array} \right)
\left( \begin{array}{ccc} \overline{c}_{13} & 0 & \overline{s}_{13} \\
0 & 1 & 0 \\ - \overline{s}_{13} & 0 & \overline{c}_{13} \end{array}
\right) \left( \begin{array}{ccc} \overline{c}_{12} & \overline{s}_{12} & 
0 \\ - \overline{s}_{12} & \overline{c}_{12} & 0 \\ 0 & 0 & 1 
\end{array} \right),
\end{equation}

\noindent
where $\overline{s}_{ij} \equiv \sin \overline{\theta}_{ij}$, and so on.
One can write $U_{\nu}$ in a similar way, with the corresponding
angles being denoted $\tilde{\theta}_{ij}$. (Henceforth, a bar over a  
quantity means that it comes from the charged lepton sector, while a tilde 
means that it comes from the neutrino sector.)
Consequently, if we assume
all quantities are real
the MNS matrix can be written 

\begin{equation}
U_{MNS} = \left( \begin{array}{ccc} 
\overline{c}_{13} \overline{c}_{12} & - \overline{s}_{12} &
-\overline{s}_{13} \overline{c}_{12} \\
- \overline{c}_{13} \overline{s}_{12} & \overline{c}_{12} &
- \overline{s}_{13} \overline{s}_{12} \\ \overline{s}_{13} & 0
& \overline{c}_{13} \end{array} \right) \left( \begin{array}{ccc}
1 & 0 & 0 \\ 0 & c_{23} & s_{23} \\ 0 & - s_{23} & c_{23} \end{array}
\right) \left( \begin{array}{ccc} \tilde{c}_{13} \tilde{c}_{12} &
\tilde{c}_{13} \tilde{s}_{12} & \tilde{s}_{13} \\
- \tilde{s}_{12} & \tilde{c}_{12} & 0 \\
- \tilde{s}_{13} \tilde{c}_{12} & - \tilde{s}_{13} \tilde{s}_{12} &
\tilde{c}_{13} \end{array} \right),
\end{equation}

\noindent
where $s_{23} \equiv \sin 
(\tilde{\theta}_{23} - \overline{\theta}_{23})$. What makes these expressions
useful is that for hierarchical mass matrices, and most of the
mass matrices that we shall have to deal with, the angles
$\overline{\theta}_{ij}$ and $\tilde{\theta}_{ij}$ are given
with sufficient accuracy very simply in terms of ratios of elements of the
mass matrices. Equation (25) tells us that

\begin{equation}
\begin{array}{ccl}
U_{e2} & = & \overline{c}_{13} \overline{c}_{12} \tilde{c}_{13} 
\tilde{s}_{12} - \overline{s}_{12} c_{23} \tilde{c}_{12} +
\overline{s}_{12} s_{23} \tilde{s}_{13} \tilde{s}_{12} 
+ \overline{s}_{13} \overline{c}_{12} s_{23} \tilde{c}_{12}
+ \overline{s}_{13} \overline{c}_{12} c_{23} \tilde{s}_{13}
\tilde{s}_{12} \\ & & \\
U_{e3} & = & - \overline{s}_{12} s_{23} \tilde{c}_{13} 
- \overline{s}_{13} \overline{c}_{12} c_{23} \tilde{c}_{13}
+ \overline{c}_{13} \overline{c}_{12} \tilde{s}_{13} \\ & & \\
U_{\mu 3} & = & \overline{c}_{12} s_{23} \tilde{c}_{13} + 
\overline{c}_{13} \overline{s}_{12} \tilde{s}_{13} 
- \overline{s}_{13} \overline{s}_{12} c_{23} \tilde{c}_{13}.
\end{array}
\end{equation}

\noindent
As we shall now see, for the practically interesting cases these 
expressions can be greatly simplified due to the smallness of
certain angles.
First let us consider the form in Eq. (2). From Eq. (2) one sees
immediately that $\tilde{s}_{23} \cong s$. From Eq. (3)
one sees that $\tilde{s}_{13} \cong (c m_{13} + s m_{12})/m_3$
and $\tilde{s}_{12} \cong (c m_{12} - s m_{13})/m_2$, where
$m_2$ and $m_3$ are the second and third eigenvalues of $M_{\nu}$.
($m_3 \cong M$ and $m_2 = O(\delta) M$.) Since $s, c \sim 1$,
one expects that $c m_{13} + s m_{12}
\sim c m_{12} - s m_{13}$, unless there is fine-tuning. Consequently,
one expects that $\tilde{s}_{13} \sim (m_2/m_3) \tilde{s}_{12} \ll
\tilde{s}_{12}$. 

The same is true in virtually all models for the charged lepton sector,
i.e. $\overline{s}_{13} \sim (m_{\mu}/m_{\tau}) \overline{s}_{12} 
\ll \overline{s}_{12}$. It is also usually true in models 
(except the flavor democracy models) that
if the 1-2 mixing is large it is due to
the angle $\tilde{\theta}_{12}$ being large 
rather than the angle $\overline{\theta}_{12}$.
In other words, $\tilde{s}_{12}$ may be large or small depending on
which solution to the solar neutrino problem is assumed, but 
$\overline{s}_{12}$ is small in almost all models (except the flavor 
democracy ones), 
implying that $\overline{s}_{13}$ is even smaller and in fact
negligible. However, $\tilde{s}_{13}$ 
can be significant if $\tilde{s}_{12} \sim 1$. 
These facts allow one to write

\begin{equation}
\begin{array}{ccl}
U_{e2} & \cong & \tilde{s}_{12} -
\overline{s}_{12} c_{23} \tilde{c}_{12} \\ & & \\
U_{e3} & \cong & - \overline{s}_{12} s_{23} + \tilde{s}_{13} \\ & & \\
U_{\mu 3} & \cong & s_{23} + \overline{s}_{12} \tilde{s}_{13} 
\end{array} 
\end{equation}

Now let us turn to models of class I(1) that have $M_{\nu}$ of the form
given in Eq. (2). There are two cases to consider: either large- or
small-angle solution to the solar neutrino problem. If small-angle
then one has that $U_{e2} \sim 0.05$, and therefore, barring
accidental cancellations, $\overline{s}_{12}, \tilde{s}_{12} 
\stackrel{_<}{_\sim} 0.05$. Thus $\tilde{s}_{13} \ll 0.05$ and
the formulas can be simplified to $U_{e3} \cong - \overline{s}_{12}
s_{23}$ and $U_{e2} \cong - \overline{s}_{12} c_{23} + \tilde{s}_{12}$.
If the solar neutrino angle is predominantly from the charged
lepton sector, i.e. $\overline{s}_{12} \gg \tilde{s}_{12}$, then
one has the predictions that $U_{e2} \cong - \overline{s}_{12} c_{23}$
and $U_{e3} \cong - \overline{s}_{12} s_{23}$, and therefore
$U_{e3} \cong U_{e2} \tan \theta_{23} \cong U_{e2} \tan \theta_{atm}$.
In other words, we have the mixing pattern $\gamma$. It is known 
experimentally that $c_{23} \approx 0.7$ and that (for small-angle MSW)
$U_{e2} \approx 0.05$, and so these relations imply that $|\overline{s}_{12}|
\cong |U_{e2}/c_{23} | \approx 0.07$. It is quite interesting that
this is numerically close to $\sqrt{m_e/m_{\mu}}$. The relation
$\overline{s}_{12} \cong \sqrt{m_e/m_{\mu}}$ is what would be
obtained in models where the 1-2 block of the charged-lepton mass matrix
has the Weinberg-Wilczek-Zee-Fritzsch form.$^{54}$ In such models one can
get the fairly sharp predictions for both $U_{e2}$ and $U_{e3}$
that we call pattern $\gamma'$. The very interesting point that
$U_{e2} = \sqrt{m_e/m_{\mu}} \cos \theta_{23}$ can arise in a simple way
and that it 
gives a good fit for the small-angle MSW solution was first emphasized 
in Ref. 29. One of the models that gives the pattern $\gamma'$ predictions
is the small-angle case of the model of Refs. 44 and 53.

The other possibility in the small solar angle case is that the solar angle 
comes predominantly from the neutrino sector, i.e. $\tilde{s}_{12}
\gg \overline{s}_{12}$. Then it is apparent that one would have
$U_{e3} \ll U_{e2}$, in other words what we called mixing pattern
$\delta$. Of course, one could have $\overline{s}_{12} \cong 
\tilde{s}_{12}$, but such a coincidence is not what one would typically
expect.

Next let us consider models of class I(1) with $M_{\nu}$ of the form
given in Eq. (2) but with large-angle solar solution. In that case,
as noted, in virtually all published models the large solar angle
comes from the neutrino sector. Thus $\tilde{s}_{12} \sim 1$ and
$\overline{s}_{12} \ll 1$. One then expects, as seen above, that
$\tilde{s}_{13} \sim (m_2/m_3) \tilde{s}_{12} \sim m_2/m_3$,
which for hierarchical models is $m_{\nu_{\mu}}/m_{\nu_{\tau}}
\cong (\delta m^2_{12}/\delta m^2_{23})^{1/2}$. For large-angle MSW
solution to the solar neutrino problem this gives $\tilde{s}_{13}
\sim 0.05$.
One typically finds in most models that 
$\overline{s}_{12} \sim \sqrt{m_e/m_{\mu}} \cong 0.07$. Thus the
two terms in the expression $U_{e3} \cong - \overline{s}_{12} s_{23}
+ \tilde{s}_{13}$ are typically of the same order but
not sharply predicted. Consequently, all one can say
is that $U_{e3} = O(m_{\nu_{\mu}}/m_{\nu_{\tau}})$ or
$\sqrt{m_e/m_{\mu}}$. In other words, one has what we called 
mixing pattern $\alpha$. Similar results follow for the vacuum
oscillation solution with hierarchical neutrino masses. In that case,
however, $\delta m^2_{12}$ is much smaller, so that one has $\tilde{s}_{13}
\sim 10^{-4}$, which is negligible. Therefore $U_{e3}$ comes from the
single term $- \overline{s}_{12} s_{23}$. In most models one has
no sharp prediction for this, and therefore the mixing pattern is
again $\alpha$. However, in some models having the WWZF form for the 
1-2 block of $L$ it is predicted that 
$\overline{s}_{12} \cong \sqrt{m_e/m_{\mu}}$, in which case the
mixing pattern is $\alpha'$. (A good example of a model 
with pattern $\alpha'$ is the
large-angle version of the model in Refs. 44 and 53. This version is
discussed in Refs. 48 and 55.)

So far we have been considering models of class I(1) in which the
matrix $M_{\nu}$ has the form given in Eq. (2). Now let us consider
the form given in Eq. (6). This form only gives large-angle solutions
to the solar neutrino problem.
It is apparent by inspection of Eq. (6) that
$\delta m^2_{23} \cong M^2$, while $\delta m^2_{12} \sim m_{ij} M$.
(More precisely, it turns out that $\delta m^2_{12} \cong
2(m_{11} + c^2 m_{22} + 2sc m_{23} + s^2 m_{33}) M$.)
Thus typically $m_{ij}/M \sim \delta m^2_{12}/\delta m^2_{23}
\sim 10^{-3}$ or $10^{-7}$ for large-angle MSW and vacuum oscillation
solutions respectively. It is straightforward to show 
that Eq. (6) gives $\tilde{s}_{13} \cong (- sc m_{22} + (c^2 - s^2) m_{23}
+ sc m_{33})/M$. Consequently, unless there is some artificial tuning
of the $m_{ij}$ one can conclude that $\tilde{s}_{13} \stackrel{_<}{_\sim}
10^{-3}$ and hence negligible. Therefore, $U_{e3} \cong - \overline{s}_{12}
s_{23}$. Generally, this means mixing pattern $\alpha$, but 
where $\overline{s}_{12}$ is predicted to be $\sqrt{m_e/
m_{\mu}}$ one has pattern $\alpha'$.

This brings us to models of class I(2). As can be seen from Eqs. (13), (15), 
and (18) most models of this class that do not involve fine-tuning
seem to yield the form for $M_{\nu}$ given in Eq. (2). These
models give the same results for $U_{e3}$
as do class I(1) models that have the form in Eq. (2). The same is also
effectively true for most models of class II(2), namely the models with
lopsided $L$. It is true that in class II(2) models the large 2-3
mixing comes from the charged lepton sector rather than from $M_{\nu}$.
However, as can be seen
from Eq. (25) it does not much matter in computing the MNS matrix
where the 2-3 mixing originates. In class II(2) models, if the 2-3 block
of the charged-lepton mass matrix $L$ is diagonalized, 
the matrix $M_{\nu}$ generally goes over to the form in Eq. (2). 
(This will be the case if the neutrino masses have the hierarchy
$m_3 \gg m_2 \gg m_1$, as typically is the case in class II(2).) 

Let us consider, finally, the models of class II(1). 
Almost all published models
of this class are of the ``flavor democracy" type, as we have seen.
Up to now we have analyzed predictions for $U_{e3}$ using the forms
given in Eqs. (24) and (25). However, these forms are convenient
when the mass matrices have a hierarchy among their elements, which
is not the case for the flavor democratic form, Eq. (19).
Therefore we shall analyze the flavor democracy models in a different
way.

In flavor democracy models, it is assumed that the lepton mass matrices
have the following forms

\begin{equation}
\begin{array}{lcl}
L & = & M_{FD} + \Delta L, \\ & & \\
M_{\nu} & = & m_{\nu} I + \Delta M_{\nu}, \end{array} 
\end{equation}

\noindent
where $M_{FD}$ is the form in Eq. (19), $I$ is the identity matrix,
and $\Delta L$ and $\Delta M_{\nu}$ are small corrections that break
the flavor permutation symmetries of the model (generally $S_3
\times S_3$). In Ref. 38 the parameter $m_{\nu}$ vanishes and 
$M_{\nu} = \Delta M_{\nu}$ has a hierarchical form, thus giving
$m_3 \ll m_2 \ll m_1$ for the three neutrino masses. But the more
usual assumption is that $m_{\nu} \neq 0$, giving $m_3 \cong
m_2 \cong m_1$. However, there is still assumed to be a hierarchy
in $\Delta M_{\nu}$ so as to get $\delta m^2_{12} \ll \delta m^2_{23}$.

The first step in diagonalizing $L$ is to transform it by the
orthogonal matrix $U_{FD}$ given in Eq. (20).

\begin{equation}
\begin{array}{lcl}
L' \equiv U_{FD}^{\dag} L U_{FD} & = & U_{FD}^{\dag} (M_{FD} + 
\Delta L) U_{FD} \\ & & \\
& = & \left( \begin{array}{ccc} 0 & 0 & 0 \\
0 & 0 & 0 \\ 0 & 0 & 3 \end{array} \right) m_{\ell} + \Delta L'.
\end{array}
\end{equation}

\noindent
Define

\begin{equation}
(\Delta L)_{ij} \equiv \delta_{ij} m_{\ell},
\;\;\; 
(\Delta L')_{ij} \equiv \delta'_{ij} m_{\ell},
\end{equation}

\noindent
Then the $\delta'_{ij}$ are given by

\begin{equation}
\begin{array}{ccl}
\delta'_{11} & = & \frac{1}{2} ( \delta_{11} + \delta_{22} - 
\delta_{12} - \delta_{21}), \\ 
\delta'_{22} & = & \frac{1}{6} ( \delta_{11} + \delta_{22} +
\delta_{12} + \delta_{21}) \\ & & - \frac{1}{3} ( \delta_{13} +
\delta_{31} + \delta_{23} + \delta_{32}) + \frac{2}{3} \delta_{33}, \\
\delta'_{33} & = & \frac{1}{3} \Sigma_{ij} \delta_{ij}, \\
\delta'_{12} & = & \frac{1}{2 \sqrt{3}} ( \delta_{11} - \delta_{22}
+ \delta_{12} - \delta_{21} - 2 \delta_{13} + 2 \delta_{23} ), \\
\delta'_{13} & = & \frac{1}{\sqrt{6}} ( \delta_{11} - \delta_{22}
+ \delta_{12} - \delta_{21} + \delta_{13} - \delta_{23}), \\
\delta'_{23} & = & \frac{1}{3 \sqrt{2}} \Sigma_i ( \delta_{1i} +
\delta_{2i} - 2 \delta_{3i}). 
\end{array}
\end{equation}

The next step in the diagonalization is to rotate away the 13, 31, 23, 
and 32 elements of $L'$ as follows

\begin{equation}
\begin{array}{ccl}
L^{\prime \prime} & = & \left( \begin{array}{ccc}
1 & 0 & - \delta'_{13}/3 \\ 0 & 1 & - \delta'_{23}/3 \\
\delta'_{13}/3 & \delta'_{23}/3 & 1 \end{array} \right) L'
\left( \begin{array}{ccc}
1 & 0 &  \delta'_{31}/3 \\ 0 & 1 &  \delta'_{32}/3 \\
- \delta'_{31}/3 & - \delta'_{32}/3 & 1 \end{array} \right) \\
& & \\
& \cong & \left( \begin{array}{ccc}
\delta'_{11} & \delta'_{12} & 0 \\ \delta'_{21} & \delta'_{22} & 0 \\
0 & 0 & 3 \end{array} \right) m_{\ell}.
\end{array}
\end{equation}

Finally the 1-2 block of $L^{\prime \prime}$ is diagonalized

\begin{equation}
L_{diag} = \left( \begin{array}{ccc} \cos \theta' & - \sin \theta' & 
0 \\ \sin \theta' & \cos \theta' & 0 \\ 0 & 0 & 1 \end{array} \right)
L^{\prime \prime} \left( \begin{array}{ccc}
\cos \theta_{\ell} & \sin \theta_{\ell} & 0 \\
- \sin \theta_{\ell} & \cos \theta_{\ell} & 0 \\ 0 & 0 & 1 \end{array}
\right), 
\end{equation}

\noindent
where

\begin{equation}
\tan 2 \theta_{\ell} = \frac{2( \delta'_{11} \delta'_{12}
+ \delta'_{21} \delta'_{22})}{(\delta^{\prime 2}_{22} +
\delta^{\prime 2}_{12} - \delta^{\prime 2}_{21} -
\delta^{\prime 2}_{11})}.
\end{equation}

\noindent
As emphasized in Ref. 56,
there is no reason {\it a priori} for this angle to be small, a
point to which we shall return presently.

Altogether, then, the matrix $U_L$ that diagonalizes $L^{\dag} L$
is given by

\begin{equation}
U_L = \left( \begin{array}{ccc}
\frac{1}{\sqrt{2}} &  \frac{1}{\sqrt{6}} & \frac{1}{\sqrt{3}} \\ 
-\frac{1}{\sqrt{2}}
& \frac{1}{\sqrt{6}} & \frac{1}{\sqrt{3}} \\ 0 &
-\frac{2}{\sqrt{6}} & \frac{1}{\sqrt{3}}  \end{array} \right)
\left( \begin{array}{ccc}
1 & 0 &  \delta'_{31}/3 \\ 0 & 1 &  \delta'_{32}/3 \\
- \delta'_{31}/3 & - \delta'_{32}/3 & 1 \end{array} \right)
\left( \begin{array}{ccc}
\cos \theta_{\ell} & \sin \theta_{\ell} & 0 \\
- \sin \theta_{\ell} & \cos \theta_{\ell} & 0 \\ 0 & 0 & 1 \end{array}
\right).
\end{equation}

\noindent
The usual assumption is that $M_{\nu}$ is nearly diagonal, so that
$U_{\nu} \cong I$ and the MNS matrix is given by $U_{MNS} 
= U_L^{\dag} U_{\nu} \cong U_L^{\dag}$. From Eq. (35) one has then 

\begin{equation}
\begin{array}{ccl}
U_{\mu 3} & \cong & - 2 \cos \theta_{\ell}/\sqrt{6} + O(\delta), \\
& & \\
U_{e2} & \cong & - \cos \theta_{\ell}/\sqrt{2} - \sin \theta_{\ell}/
\sqrt{6} + O(\delta), \\
& & \\
U_{e3} & \cong & 2 \sin \theta_{\ell}/\sqrt{6} - (\sin 
\theta_{\ell} \delta'_{32} - \cos \theta_{\ell} \delta'_{31})/3\sqrt{3}.
\end{array}
\end{equation}

\noindent
Since the angle $\theta_{\ell}$ is very sensitive to the nine parameters
$\delta_{ij}$ and has no reason {\it a priori} to be small, as is
apparent from Eq. (34), it might seem that the flavor democracy idea
has no predictivity as far as the MNS matrix elements are concerned.
However, {\it a posteriori} we do know that the CKM angles are small,
and that strongly suggests that $\theta_{\ell}$ is small. The point is that
if a large angle $\theta_{\ell}$ were required in the diagonalization of
$L$, one would typically expect to find that large
angles $\theta_u$ and $\theta_d$ were required in the diagonalization of
$U$ and $D$ as well. Unless there were a conspiracy and $\theta_u \cong
\theta_d$, large CKM angles would result. Under the assumption that
$\Delta L$ has the same form (with different values of parameters) as 
$\Delta U$ and $\Delta D$, one can conclude that $\theta_{\ell} \ll 1$.

There are many possible forms for $\Delta L$ that give vanishing 
$\theta_{\ell}$. If such a form is chosen, then one has 

\begin{equation}
\begin{array}{ccl}
U_{\mu 3} & = & - 2/\sqrt{6} + O(\delta), \\ & & \\
U_{e2} & = & - 1/\sqrt{2} + O(\delta), \\ & & \\
U_{e3} & = & O(\delta).
\end{array}
\end{equation}

\noindent
The exact value is evidently dependent on the scheme of symmetry
breaking. However, since the parameters $\delta_{ij}$ are involved
in generating the interfamily hierarchy of of charged lepton masses,
one expects that $U_{e3}$ will closely related to small lepton mass
ratios. In fact, this is the case, and typically one finds that
$U_{e3} \sim \sqrt{m_e/m_{\mu}}$, in other words pattern $\alpha$.
In a popular scheme of symmetry
breaking,$^{57,58}$ for instance, $| U_{e3}| \cong \frac{2}{\sqrt{6}} 
\sqrt{m_e/m_{\mu}}$, which we have called pattern $\alpha^{\prime \prime}$. 
However, there are also
schemes of symmetry breaking$^{57}$ where $U_{e3} = 0$, which we called
pattern $\beta$. 

In conclusion, we see that there are a few patterns of neutrino mixing 
that tend to arise in the great majority of published models. And although
there is not a one-to-one correspondence between the type of model
and the value of $U_{e3}$, it is clear that knowledge of $U_{e3}$
will give great insight into the possible underlying mechanisms 
that are responsible for neutrino mixing.$^{59}$

\section*{References}

\begin{enumerate}
\item J.W.F. Valle, 
hep-ph/9911224; S.M. Bilenky,  
Lectures at the 1999 European School of High Energy
Physics, Casta Papiernicka, Slovakia, Aug. 22-Sept. 4, 1999
hep-ph/0001311.
\item M.C. Gonzalez-Garcia, P.C. de Holanda, C. Pe\~{n}a-Garay, and
J.C.W. Valle, hep-ph/9906469
\item V. Barger and K. Whisnant, hep-ph/9903262
\item M.C. Gonzalez-Garcia, talk at International Workshop on Particles
in Astrophysics and Cosmology: From Theory to Observation, Valencia,
Spain, May 3-8, 1999.
\item M. Gell-Mann, P. Ramond, and R. Slansky, in {\it Supergravity,
Proc. Supergravity Workshop at Stony Brook}, ed. P. Van Nieuwenhuizen
and D.Z. Freedman (North-Holland, Amsterdam (1979)); T. Yanagida,
{\it Proc. Workshop on unified theory and the baryon number of the
universe}, ed. O. Sawada and A. Sugamoto (KEK, 1979).
\item A. Zee, {\it Phys. Lett.} {\bf B93}, 389 (1980); 
{\it Phys. Lett.} {\bf B161}, 141 (1985).
\item C. Froggatt and H.B. Nielson, {\it Nucl. Phys.} {\bf B147}, 277 (1979).
\item J.A. Harvey, D.B. Reiss, and P. Ramond, {\it Nucl. Phys.}
{\bf B199}, 223 (1982).
\item Z. Maki, M. Nakagawa, and S. Sakata, {\it Prog. Theor. Phys.} 
{\bf 28}, 870 (1962).
\item R.N. Mohapatra and S. Nussinov, Phys. Rev. {\bf D60}, 013002 (1999)
(hep-ph/9809415).
\item C.D. Froggatt, M. Gibson, and H.D. Nielson, {\it Phys. Lett.}
{\bf B446}, 256 (1999) (hep-ph/9811265). 
\item A.S. Joshipura, hep-ph/9808261; A.S. Joshipura and 
S.D. Rindani, hep-ph/9811252; R.N. Mohapatra, A. Perez-Lorenzana,
C.A. deS. Pires, {\it Phys. Lett.} {\bf B474}, 355 (2000)
(hep-ph/9911395).
\item C. Jarlskog, M. Matsuda, S. Skadhauge, and M. Tanimoto,
{\it Phys. Lett.} {\bf B449}, 240 (1999) (hep-ph/9812282).
\item E. Ma, {\it Phys. Lett.} {\bf B442}, 238 (1998) (hep-ph/9807386);
K. Cheung and O.C.W. Kong, hep-ph/9912238.
\item P. Frampton and S. Glashow, {\it Phys. Lett.} {\bf B461}, 95 (1999)
(hep-ph/9906375);
A.S. Joshipura and S.D. Rindani, {\it Phys. Lett.} {\bf B464}, 239
(1999) (hep-ph/9907390).
\item M. Drees, S. Pakvasa, X. Tata, T. terVeldhuis, 
{\it Phys. Rev.} {\bf D57} 5335 (1998) (hep-ph/9712392).
\item E.J. Chun, S.K. Kang, C.W. Kim, and U.W. Lee, {\it Nucl.
Phys.} {\bf B544}, 89 (1999) (hep-ph/9907327);
A.S. Joshipura and S.K. Vempati, {\it Phys. Rev.} {\bf D60},
095009 (1999) (hep-ph/9808232); 
B. Mukhopadhyaya, S. Roy, and F. Vissani, {\it Phys.
Lett.} {\bf B443}, 191 (1998) (hep-ph/9808265); O.C.W. Kong, hep-ph/9808304;
K. Choi, E.J. Chun, and K. Hwang, {\it Phys. Rev.} {\bf D60}, 031301
(1999) (hep-ph/9811363); D.E. Kaplan and A.E. Nelson, {\it JHEP} 0001:033
(2000) (hep-ph/9901254); A.S. Joshipura and S.K. Vempati, 
{\it Phys. Rev.} {\bf D60}, 111303 (1999) (hep-ph/9903435); 
J.C. Romao, M.A. Diaz, M. Hirsch, W. Porod, and J.W.F. Valle,
hep-ph/9907499; O. Haug, J.D. Vergados, A. Faessler, and S. Kovalenko,
hep-ph/9909318; E.J. Chun and S.K. Kang, {\it Phys. Rev.} 
{\bf D61}, 075012 (2000) (hep-ph/9909429).
\item K. Fukuura, T. Miura, E. Takasugi, and M. Yoshimura, 
Osaka Univ. preprint, OU-HET-326 (hep-ph/9909415).
\item G.K. Leontaris and J. Rizos, CERN-TH-99-268 (hep-ph/9909206);
W. B\"{u}chmuller and T. Yanagida, {\it Phys. Lett.} {\bf B445}, 399 (1999)
(hep-ph/9810308); C.K. Chua, X.G. He, and W.Y. Hwang, hep-ph/9905340;
A. Ghosal, hep-ph/9905470; J.E. Kim and J.S. Lee, hep-ph/9907452;
U. Mahanta, hep-ph/9909518.
\item S.L. King, {\it Phys. Lett.} {\bf B439}, 350 (1998) (hep-ph/9806440);
S. Davidson and S.L. King, {\it Phys. Lett.} {\bf B445}, 191 (1998) 
(hep-ph 9808296); E. Ma and D.P Roy, {\it Phys. Rev.} {\bf D59}, 097702
(1999) (hep-ph/9811266); Q. Shafi and Z. Tavartkiladze, {\it Phys.
Lett.} {\bf B451}, 129 (1999) (hep-ph/9901243); S. King, {\it Nucl.
Phys.} {\bf B562}, 57 (1999) (hep-ph/9904210);
W. Grimus and H. Neufeld, hep-ph/9911465.
\item R. Barbieri, L.J. Hall, D. Smith, A. Strumia, and N. Weiner,
{\it JHEP}, 9812:017 (1998) (hep-ph/9807235);
R. Barbieri, L.J. Hall, and A. Strumia, {\it Phys. Lett.} 
{\bf B445}, 407 (1999) (hep-ph/9808333); Y. Grossman, Y. Nir, and Y. Shadmi,
{\it JHEP} 9810:007 (1998) (hep-ph/9808355); C.D. Froggatt, M. Gibson,
and H.D. Nielson, {\it Phys. Lett.} {\bf B446}, 256 (1999) (hep-ph/9811265).
\item C.H. Albright and S. Nandi, {\it Phys. Rev.} {\bf D53}, 2699
(1996) (hep-ph/9507376); H. Nishiura, K. Matsuda, and T. Fukuyama, 
{\it Phys. Rev.} {\bf D60}, 013006 (1999) (hep-ph/9902385); K.S. Babu,
B. Dutta, and R.N. Mohapatra, {\it Phys. Lett.} {\bf B458}, 93 (1999)
(hep-ph/9904366).
\item M. Jezabek and Y. Sumino, 
{\it Phys. Lett.} {\bf B440}, 327 (1998) (hep-ph/9807310); 
G. Altarelli and F. Feruglio, 
{\it Phys. Lett.} {\bf B439}, 112 (1998)
(hep-ph/9807353). G. Altarelli, F. Feruglio, and I. Masina,
hep-ph/9907532.
\item B. Stech, {\it Phys. Lett.} {\bf B465}, 219 (1999) 
(hep-ph/9905440);
R. Derm\'{i}\u{s}ek and S. Raby, hep-ph/9911275; 
A. Aranda, C.D. Carone, and R.F. Lebed, hep-ph/0002044.
\item G.K. Leontaris, S. Lola, C. Scheich, J.D. Vergados, {\it Phys.
Rev.} {\bf D53}, 6381 (1996) (hep-ph/9509351); Y. Koide, {\it Mod.
Phys. Lett.} {\bf A11}, 2849 (1996) (hep-ph/9603376); P. Binetruy, S.
Lavignac, S. Petcov, and P. Ramond, {\it Nucl. Phys.} {\bf B496},
3 (1997) (hep-ph/9610481); B.C. Allanach, {\it Phys. Lett.} {\bf B450},
182 (1999) (hep-ph/9806294); G. Eyal, {\it Phys. Lett.} {\bf B441},
191 (1998) (hep-ph/9807308); S. Lola and J.D. Vergados, {\it Prog.
Part. Nucl. Phys.} {\bf 40}, 71 (1998) (hep-ph/9808269).
\item G. Costa and E. Lunghi, {\it Nuov. Cim.} {\bf 110A}, 549 (1997)
(hep-ph/9709271).
\item M. Jezabek and Y. Sumino, 
{\it Phys. Lett.} {\bf B440}, 327 (1998) (hep-ph/9807310).
\item G. Altarelli and F. Feruglio, {\it Phys. Lett.} {\bf B439}, 112 (1998) 
(hep-ph/9807353); E.Kh. Akhmedov, G.C. Branco, and M.N. Rebelo,
hep-ph/9911364.
\item M. Bando, T. Kugo, and K. Yoshioki, 
{\it Phys. Rev. Lett.} {\bf 80}, 3004 (1998) (hep-ph/9710417).
\item M. Abud, F. Buccella, D. Falcone, G. Ricciardi, and
F. Tramontano, DSF-T-99-36 (hep-ph/9911238). 
\item A.K. Ray and S. Sarkar, {\it Phys. Rev.} {\bf D61}, 035007 (2000)
(hep-ph/9908294).
\item J. Hashida, T. Morizumi, and A. Purwanto, hep-ph/9909208.
\item K. Oda, E. Takasugi, M. Tanaka, and M. Yoshimura, {\it Phys.
Rev.} {\bf D59}, 055001 (1999) (hep-ph/9808241).
\item Q. Shafi and Z. Tavartkiladze, BA-99-39 (hep-ph/9905202);
D.P. Roy, Talk at 6th Topical Seminar on Neutrino and AstroParticle Physics,
San Miniato, Italy, 17-21 May 1999 (hep-ph/9908262).
\item G. Altarelli, F. Feruglio, and I. Masina, hep-ph/9907532.
See also R. Barbieri, P. Creminelli, and A. Romanino, {\it Nucl. Phys.}
{\bf B559}, 17 (1999) (hep-ph/9903460).
\item E. Malkawi, {\it Phys. Rev.} {\bf D61}, 013006 (2000) (hep-ph/9810542);
Y.L. Wu, {\it Eur. Phys. J.} {\bf C10}, 491 (1999) (hep-ph/9901245).
\item For a review see H. Fritzsch, Talk at Ringberg Euroconference on
New Trends in Neutrino Physics, Ringberg, Ger. 1998 (hep-ph/9807234),
and references therein.
\item H. Fritzsch and Z.Z. Xing, {\it Phys. Lett.} {\bf B372}, 265
(1996) (9509389).
\item M. Fukugita, M. Tanimoto, and T. Yanagida, 
{\it Phys. Rev.} {\bf D57}, 4429 (1998) (hep-ph/9709388);
M. Tanimoto, {\it Phys. Rev.} {\bf D59}, 017304 (1999) (hep-ph/9807283);
H. Fritzsch and Z.Z. Xing, 
{\it Phys. Lett.} {\bf B440}, 313 (1998) (hep-ph/9808272);
R.N. Mohapatra and S. Nussinov, {\it Phys. Lett.} {\bf B441}, 299 (1998)
(hep-ph/9808301); M. Fukugita, M. Tanimoto, and T. Yanagida, {\it Phys.
Rev.} {\bf D59}, 113016 (1999) (hep-ph/9809554); S.K. Kang and C.S. Kim, 
{\it Phys. Rev.} {\bf D59}, 091302 (1999) (hep-ph/9811379);
M. Tanimoto, T. Watari, and T. Yanagida, {\it Phys. Lett.}
{\bf B461}, 345 (1999) (hep-ph/9904338);
M. Tanimoto, hep-ph/0001306. For an excellent review of flavor democracy
schemes of neutrino mass mixing see H. Fritzsch and Z.Z. Xing,
hep-ph/9912358.
\item G.C. Branco, M.N. Rebelo, and J.I. Silva-Marcos, {\it Phys.
Lett.} {\bf B428}, 136 (1998) (hep-ph/9802340); I.S. Sogami,
H. Tanaka, and T. Shinohara, {\it Prog. Theor. Phys.} {\bf 101}, 707
(1999) (hep-ph/9807449); G.C. Branco, M.N. Rebelo, and J.I. Silva-Marcos,
hep-ph/9906368.
\item K.S. Babu and S.M. Barr, 
{\it Phys. Lett.} {\bf B381}, 202 (1996) (hep-ph/9511446).
\item S.M. Barr, 
{\it Phys. Rev.} {\bf D55}, 1659 (1997) (hep-ph/9607419). 
\item J. Sato and T. Yanagida, {\it Phys. Lett.} {\bf B430}, 127
(1998) (hep-ph/9710516).
\item C.H. Albright, K.S. Babu, and S.M. Barr, 
{\it Phys. Rev. Lett.} {\bf 81}, 1167 (1998) (hep-ph/9802314).
\item N. Irges, S. Lavignac, and P. Ramond, {\it Phys. Rev.} {\bf D58},
035003 (1998) (hep-ph/9802334); J.K. Elwood, N. Irges, and P. Ramond,
{\it Phys. Rev. Lett.} {\bf 81}, 5064 (1998) (hep-ph/9807228).
\item Y. Nomura and T. Yanagida, 
{\it Phys. Rev.} {\bf D59}, 017303 (1999) (hep-ph/9807325);
N. Haba, {\it Phys. Rev.} {\bf D59}, 035011 (1999) (hep-ph/9807552);
G. Altarelli and F. Feruglio, {\it JHEP} 9811:021 (1998) (hep-ph/9809596);
Z. Berezhiani and A. Rossi, {\it JHEP} 9903:002 (1999) (hep-ph/9811447);
K. Hagiwara and N. Okamura, 
{\it Nucl. Phys.} {\bf B548}, 60 (1999) (hep-ph/9811495);
G. Altarelli and F. Feruglio, 
{\it Phys. Lett.} {\bf B451}, 388 (1999) (hep-ph/9812475);
K.S. Babu, J. Pati, and F. Wilczek, (hep-ph/9812538);
M. Bando and T. Kugo, {\it Prog. Theor. Phys.} {\bf 101}, 1313 (1999) 
(hep-ph/9902204); Y. Nir and Y. Shadmi, {\it JHEP} 9905:023 (1999)
(hep-ph/9902293); Y. Nomura and T. Sugimoto, hep-ph/9903334;
K.I. Izawa, K. Kurosawa, Y. Nomura, and T. Yanagida, 
{\it Phys. Rev.} {\bf D60}, 115016 (1999) (hep-ph/9904303);
Q. Shafi and Z. Tavartkiladze, BA-99-63 (hep-ph/9910314);
P. Frampton and A. Rasin, IFP-777-UNC (hep-ph/9910522).
\item R. Barbieri, L.J. Hall, G.L. Kane, and G.G. Ross, OUTP-9901-P
(hep-ph/9901228); E. Ma, {\it Phys. Rev.} {\bf D61}, 033012
(hep-ph/9909249).
\item C.H. Albright and S.M. Barr, {\it Phys. Lett.} {\bf 461}, 218
(1999) (hep-ph/9906296).
\item C.H. Albright and S.M. Barr, {\it Phys. Rev.} {\bf D58}, 013002 (1998)
(hep-ph/9712488).
\item H. Georgi and C. Jarlskog, {\it Phys. Lett.} {\bf B86} (1979) 297.
\item S.M. Barr and S. Raby, 
{\it Phys. Rev. Lett.} {\bf 79}, 4748 (1998).
\item S. Dimopoulos and F. Wilczek, report No. NSF-ITP-82-07 (1981),
in {\it The unity of fundamental interactions} Proceedings of the 19th
Course of the International School of Subnuclear Physics, Erice, Italy, 
1981 ed. A. Zichichi (Plenum Press, New York, 1983);
K.S. Babu and S.M. Barr, 
{\it Phys. Rev.} {\bf D48}, 5354 (1993); {\it Phys. Rev.} {\bf D50}, 3529
(1994).
\item C.H. Albright and S.M. Barr, 
{\it Phys. Lett.} {\bf B452}, 287 (1999) (hep-ph/9901318).
\item S. Weinberg, {\it Trans. NY Acad. Sci.} {\bf 38}, 185 (1977);
F. Wilczek and A. Zee, {\it Phys. Lett.} {\bf B70}, 418 (1977);
H. Fritzsch, {\it Phys. Lett.} {\bf B70}, 436 (1977).
\item C.H. Albright and S.M. Barr, hep-ph/0002155.
\item M. Tanimoto, {\it Phys. Rev} {\bf D59}, 017304 (1999) (hep-ph/9807283).
\item M. Fukugita, M. Tanimoto, and T. Yanagida, {\it Phys. Rev.}
{\bf D57}, 4429 (1998) (hep-ph/9709388). 
\item H. Fritzsch and Z.Z. Xing,
{\it Phys. Lett.} {\bf B440}, 313 (1998) (hep-ph/9808272); 
H. Fritzsch and Z.Z. Xing, hep-ph/9912358.
\item E.Kh. Akhmedov, G.C. Branco, and M.N. Rebelo, hep-ph/9912205
gives an analysis of $U_{e3}$ that is different from but not inconsistent
with the one given here.

\end{enumerate}

\end{document}